\documentstyle[art10]{article}

\author{M. Temple-Raston\\
Mathematical Institute,\\
Oxford University,\\
24-29 St. Giles',\\
Oxford}
\title{The redshift in Hubble's constant
}

\begin{document}

\maketitle
\begin{abstract}
A generalisation to electrodynamics and Yang-Mills theory is presented that
permits computation of the speed of light. The model presented herewithin
indicates that the speed of light is not a universal constant. This may be
relevant to the current debate in astronomy over large values of the Hubble
constant obtained by the latest generation of ground and space-based
telescopes. An experiment is proposed based on Compton scattering.
\end{abstract}

\section{Introduction}

Astronomers have recently reported on observations for the Hubble constant
that predict an age for the universe younger than the estimated age of some
globular clusters \cite{nature1}\cite{nature2}\cite{nature3}. The obvious
tension contained in this result has been christened the `age crisis'. There
are three possibilities of course: the measurements made for the Hubble
constant are incorrect; current models for stellar evolution are incorrect;
or, there is new physics to be understood. Without further observation it is
too early to view favourably recent experiments for the Hubble constant. And
as noted by Sandage, if the results are supported by further
experimentation, it must be understood why previous estimates for the Hubble
constant based on observations of type Ia supernovae are only half as large.
It may be that models for stellar evolution need to be re-evaluated as well.

Over the course of the last fifteen years, observation has uncovered
deviations in the Hubble constant. The Hubble relation turns sharply upwards
from linearity on a redshift vs. distance plot \cite{sandage}\cite{spinrad}.
Refinements to the distance scale achieved by \cite{nature1}\cite{nature2}%
\cite{nature3} cannot of course account for the non-linearity in the Hubble
constant. Sandage has argued that the non-linearity is due to a bias in the
choice of objects chosen by astronomers for study, and that at large
distances such a bias must be filtered out in order to correctly determine
the Hubble constant. Such is the current debate.

Another hypothesis is suggested by the deviations in the Hubble constant:
the speed of light in vacuum at space-time points remote from Earth is less
than the present, terrestial speed of light. For suppose that the speed of
light in a vacuum were not an absolute constant, then the red-shift would
need to be reassessed. The redshift is given by
\begin{equation}
\label{redshift}\frac{\Delta \lambda }\lambda \cong \frac v{\tilde c},
\end{equation}
where we assume that the velocity of the emitter $v\ll \tilde c$, the speed
of light when the photon is emitted. If the speed of light at the time of
emission is smaller than the present speed of light on earth, then the
observed red-shift in (\ref{redshift}) would be greater. This would make the
Hubble constant larger, and thereby make the universe appear to be younger
than it actually is. A lower speed of light in the early universe might
therefore be helpful in understanding the purported `age crisis'. In
addition, one notes that under these circumstances the enormous powers
associated with quasars would be reduced. In the next section we present a
model that permits us to derive the speed of light in a natural way. In
section three we propose an experiment based on Compton scattering that
might detect variations in the speed of light.

\section{A model}

Let $\pi :P\rightarrow M$ be a principal $U(2)$-bundle over an oriented,
compact, connected four-manifold $M$, and denote by $E$ the associated rank
two adjoint vector bundle. ${\cal A}(P)$ is the space of connections on $P$.
Let $A,B\in {\cal A}(P)$, and introduce local coordinate charts with indices
$\mu =0,\dots ,3$ on $M$. The Lie algebra-valued connections or vector
potentials, $A_\mu $ and $B_\mu $, induce exterior covariant derivatives $%
D_\mu ^A=\partial _\mu +A_\mu $ and $D_\mu ^B=\partial _\mu +B_\mu $ on the
associated adjoint vector bundle $E$. The curvatures $H^A$ and $K^B$ are
defined by $2D_{[\mu }^AD_{\nu ]}^As=H_{\mu \nu }^As$ and $2D_{[\mu
}^BD_{\nu ]}^Bs=K_{\mu \nu }^Bs$. In this way $H^A$ and $K^B$ are two-forms
on $M$ taking values in $E$, that is $H^A,K^B\in \Lambda ^2(M,E)$. The
Lagrangian Action that forms the basis of our model is given by
\begin{equation}
\label{TFT4}{\cal L}(A,B)=\int_M<(H^A\otimes I_E)\wedge (I_E\otimes
K^B)>-\frac 12<(I_E\otimes K^B)^2>,
\end{equation}
where $H^A$ and $K^B$ are gauge field curvatures over the four-manifold, $M$%
. $I_E$ denotes the identity transformation on the adjoint bundle, $E$, and
the bundle inner product is represented by $<\ >$. The Killing-Cartan inner
product can be adopted. The inner product is normalized so that $<I_E^2>=1$.
The form of the Lagrangian (\ref{TFT4}) generalises the topological gauge
field theories studied some time ago by Horowitz \cite{horowitz}. In local
space-time coordinates and using the Killing-Cartan inner product the
Lagrangian Action can be written explicitly as
\begin{equation}
\label{Lang}
\begin{array}{ccl}
{\cal L}(A,B) & = & \int_MH_{[\mu \nu }^aK_{\lambda \rho ]}^b\
{\rm tr}(T^aI_E){\rm tr}(T^bI_E)\ d^4x \\  &  & -\frac 12\int_MK_{[\mu \nu
}^aK_{\lambda \rho ]}^b\ {\rm tr}(I_EI_E)\ {\rm tr}(T^aT^b)\ d^4x.
\end{array}
\end{equation}
The generators of the Lie algebra are denoted by $T^a$. For gauge groups
where ${\rm tr}(T^a)=0$, the Lagrangian (\ref{Lang}) reduces to the second
integral---these are the topological field theories studied by Baulieu and
Singer \cite{singer}. The variational field equations for the Lagrangian (%
\ref{TFT4}) are
\begin{equation}
\label{field}D^AK^B=0,\qquad D^BH^A=0.
\end{equation}
The field equations are clearly independent of any metric structure on $M$.
The set of solutions to (\ref{field}) is clearly not trivial, because when $%
A=B$ the field equations reduce to the Bianchi identities.

Observe that when a space-time metric is placed on $M$ (by whatever means)
and used to define the Hodge star operator,$\ ^{*}$, the topological field
equations (\ref{field}) become the source-free Yang-Mills or electrodynamic
field equations when the gauge field $B$ is chosen so that $K^B=\ ^{*}H^A$.
The topological field theory above therefore, is seen to contain Yang-Mills
theory and electromagnetism. It is for this reason that we defined a field
theory with {\it two} vector potentials, $A$ and $B$, and not one.

We turn now to the Bogomol'nyi structure. By completing the square, the
Lagrangian (\ref{TFT4}) can be rewritten as
\begin{equation}
\label{Lanbog}
\begin{array}{c}
2
{\cal L}=\int_M<(H^A\otimes I_E-I_E\otimes K^B)^2> \\ -\int_M<(H^A\otimes
I_E)^2>.
\end{array}
\end{equation}
The Lagrangian (\ref{Lanbog}) is now in Bogomol'nyi form. The Bogomol'nyi
equations arising from (\ref{Lanbog}) are
\begin{equation}
\label{bog4}H^A\otimes I_E-I_E\otimes K^B=0.
\end{equation}
By an index computation equations (\ref{bog4}) imply that $H^A$ and $K^B$
are projectively flat,
\begin{equation}
\label{bog4'}H^A=K^B=iFI,
\end{equation}
where $F$ is a real-valued two form on $M$. Clearly solutions to the
Bogomol'nyi equations (\ref{bog4}) automatically satisfy the variational
field equations (\ref{field}) when $F$ is closed.

Let $E_A$ and $E_B$ be the adjoint vector bundles equipped with $D^A$ or $%
D^B $, respectively. $E^{*}$ is the dual bundle to $E$. The curvature of the
tensor product bundle $E_A\otimes E_B^{*}$ is given by \cite{kob}
$$
\Omega _{E_A\otimes E_B^{*}}=H^A\otimes I_E-I_E\otimes K^B.
$$
In view of this, the Bogomol'nyi equations in (\ref{bog4}) are seen to be a
vanishing curvature condition on the tensor product bundle $E_A\otimes
E_B^{*}$, and implies that
\begin{equation}
\label{topcond}c_2(E\otimes E^{*})-\frac 12c_1(E\otimes
E^{*})^2=4c_2(E)-c_1(E)^2=0.
\end{equation}
The second integral in (\ref{Lanbog}) can also be written as a
characteristic class,
\begin{equation}
\label{stable}\int_M{\rm tr}(H^A\wedge H^A)=8\pi ^2(c_2(E)-\frac
12c_1(E)^2)=-8\pi ^2ch_2(E).
\end{equation}
The second Chern character is denoted by $ch_2(E)$. It is clear that under a
perturbation of the vector potentials both integrals in (\ref{Lanbog}) are
invariant, and when the Bogomol'nyi equations are satisfied the Lagrangian
is proportional to the characteristic class (\ref{stable}). Non-singular
solutions to the Bogomol'nyi equations are non-trivial and stable when (\ref
{stable}) is non-vanishing.

We have therefore identified a class of non-singular, finite-energy, stable
solutions to the variational field equations (\ref{field}). However, most
solutions to the Bogomol'nyi equations do not appear to be `particle-like'.
Thus within the class of Bogomol'nyi solutions we must now define the
`particle-like' solutions to the field equations: solutions to the
Bogomol'nyi equations (\ref{bog4'}) will be said to be `particle-like' if
together they form a phase space manifold that is non-singular, Hausdorff,
and of finite dimension. This is a tacit assumption in the particle picture,
both classically and quantum mechanically. The phase space is given by the
moduli space of solutions to the Bogomol'nyi equations (\ref{bog4'}) defined
over space-time, $M$. The phase space is generally of infinite dimension and
is not necessarily Hausdorff. In algebraic geometry, mathematicians look for
Mumford-Takemoto topological stability to ensure that moduli spaces are
Hausdorff. Kobayashi has reformulated Mumford-Takemoto stability into a
differential geometric form, known as the Einstein-Hermitian condition \cite
{kob}. As we shall see, the Einstein-Hermitian condition is equivalent to
restricting to those solutions of the Bogomol'nyi equations that are
compatible with some additional geometric structure.

To implement the Einstein-Hermitian condition we assume that space-time, $M$%
, is now a compact K\"ahler manifold of real dimension four with a K\"ahler
form $\Phi $, and that $E\rightarrow M$ is holomorphic. A Hermitian metric $%
h $ and a holomorphic structure $\bar \partial $ on $E$ give rise to a
unique connection, $A$. The associated curvature $H^A$ is of type $(1,1)$,
that is, $H^A\in \Lambda ^{1,1}(M,E)$. The mean curvature, $K$, of the
vector bundle over a K\"ahler manifold is given by \cite[p.99]{kob}%
$$
K=i\Lambda H^A,
$$
where the operator $\Lambda :\Lambda ^{n,m}\rightarrow \Lambda ^{n-1,m-1}$
is defined as the adjoint operator to $L\equiv \Phi \wedge \cdot $. A vector
bundle is said to be Einstein-Hermitian when
\begin{equation}
\label{mean}K=i\Lambda H^A=kI_E,
\end{equation}
for $k$ a real constant. When $k$ is a non-constant function on $M$, then
the vector bundle is said to be weak Einstein-Hermitian.

One cannot help but notice the similarity of equation (\ref{mean}) with the
projective flatness
\begin{equation}
\label{eqn}H^A=FI_E,
\end{equation}
of the Bogomol'nyi equations (\ref{bog4'}). For holomorphic vector bundles $%
F\in \Lambda ^{1,1}(M)$. Applying the operator $\Lambda $ to both sides of (%
\ref{eqn}), we see that a projectively flat holomorphic vector bundle over a
K\"ahler surface is weak Einstein-Hermitian:%
$$
K=i\Lambda H^A=i(\Lambda F)I_E=\varphi (x)I_E,
$$
with $\varphi (x)$ a real-valued function on $M$. In addition, it can be
shown that there exists a conformal change to the Hermitian structure $%
h\rightarrow h^{\prime }=\alpha h$ such that $(E,h^{\prime })$ satisfies the
Einstein-Hermitian condition with a constant factor $k$ \cite[p.104]{kob}. $%
k $ depends only on $c_1(E)$ and the cohomology class of $\Phi $, in
particular it is independent of the Hermitian inner product $h$. Now in the
other direction, by a theorem of L\"ubke it is known that all
Einstein-Hermitian vector bundles satisfying the topological requirement (%
\ref{topcond}) are holomorphic projectively flat \cite{lubke}. Thus, up to a
conformal change in the Hermitian structure, a holomorphic projectively flat
connection and an Einstein-Hermitian connection are equivalent.

To obtain a phase space that is topologically well-behaved we shall
therefore restrict to the set of holomorphic projectively flat connections
on a vector bundle $E$ satisfying the topological condition (\ref{topcond}),
equivalently, restrict to Einstein-Hermitian connections. We shall call the
Einstein-Hermitian Bogomol'nyi solitons `topological instantons', because of
the obvious similarity between these objects and the self-dual instantons in
Yang-Mills theory. Topological instantons are naturally parametrized by a
continuous parameter, the Einstein-Hermitian constant $k$ in (\ref{mean}).
For fixed $k$, the phase space is denoted by ${\cal M}_k$.

Now let the underlying space-time manifold be a flat K\"ahler complex
two-torus (real dimension four); we assume a space-time that is periodic in
both space and time. All complex tori admit a K\"ahler structure: a Kahler
metric $g$ and a closed Kahler form $\Phi \in \Lambda ^{1,1}(M)$ \cite{wells}%
. The complex rank two vector bundle $(E,h)\rightarrow (M,g,\Phi )$ defined
over the K\"ahler torus is assumed to have the Chern numbers $%
4c_2(E)=c_1(E)^2=-4$. $h$ is the Hermitian metric on $E$ which can be
constructed using the Killing-Cartan form. The significance behind the
choice of topology is that such a bundle satisfies the topological condition
(\ref{topcond}) necessary for the bundle to admit projectively flat
connections, and the Lagrangian (\ref{stable}) equals $-8\pi ^2$ so that
{\it stable} topological instantons might exist. For abelian varieties
existence theorems are known \cite{antony}.We shall study `diagonal' $U(2)$
topological instantons on this bundle. By `diagonal' we mean that the
Einstein-Hermitian connections $A$ and $B$ are equal ($A=B$). Diagonal
instantons are examined because there is little physical evidence to suggest
two distinct vector potentials. We take the constant $k$ in the
Einstein-Hermitian condition (\ref{mean}) to be fixed and non-zero. The
K\"ahler structure on $M$ allows us to study the phase space, ${\cal M}_k$.

As we have seen the phase space, ${\cal M}_k$, of $U(2)$ topological
instantons on the holomorphic Hermitian vector bundle $(E,h)\rightarrow M$
is equivalent to the moduli space of irreducible Einstein-Hermitian
connections ${\cal E}(E,h)/U(E,h)$ on $(E,h)$. $U(E,h)$ denotes the unitary
gauge transformations of $(E,h)$. The complex dimension of the $U(2)$
topological instanton phase space when non-empty is given by \cite{kob}:
\begin{equation}
\label{dimen}\dim {}_{{\bf C}}({\cal M}_k)=4h^{0,1}(M)-6.
\end{equation}
The K\"ahler torus has $h^{0,1}(M)=2$. Therefore if a (diagonal) topological
instanton exists, the real dimension of the phase space, ${\cal M}_k$, is
four. Note that massless particles have phase spaces of real dimension four
in (3+1) space-time. (We have skipped over K3 surfaces as models for
space-time because they have $h^{0,1}(M)=0$, thereby giving phase space a
negative dimension.)

The next order of business is to include special relativity into the theory
locally by modelling ${\cal M}_k$ with ${\cal M}_{massless}$, the phase
space for massless particles. Consider the phase space for massless
particles in ${\bf R}^3$. The (covariant) phase space is equivalent to the
space-of-motions. Massless particles in ${\bf R}^3$ move on straight lines
and at the speed of light, $c$. We may therefore parameterize the possible
motions of a massless particle by assigning to a straight line, ${\bf x}(t)$%
, in ${\bf R}^3$ a velocity vector ${\bf c}=c{\bf \hat n}$, and a position
vector ${\bf d}$ so that ${\bf x}(t)={\bf d}+t{\bf c}$. The position vector $%
{\bf d}$ is defined as the normal from the origin to the line, ${\bf x}(t)$,
equivalently, the point on the line nearest the origin. Thus the phase space
is
\begin{equation}
\label{photon}{\cal M}_{massless}\cong \{({\bf c},{\bf d})\in S_c^2\times
R^3|\ {\bf c}.{\bf d}=0\}.
\end{equation}
The phase space of the massless particle on ${\bf R}^3$ is equivalent to the
tangent bundle $TS_c^2$ where the radius of the sphere is the speed of
light. The natural metric on $TS_c^2$ is given by
\begin{equation}
\label{phomet}ds^2=f(r)dr^2+a(r)(d\psi +\cos \theta \ d\varphi
)^2+c^2(d\theta ^2+\sin {}^2\theta \,d\varphi ^2),
\end{equation}
where $(\theta ,\varphi )$ are spherical polar coordinates on the
two-sphere, and $(r=\parallel {\bf d}\parallel ,\psi )$ plane polar
coordinates on the tangent plane. The functions $f(r)$ and $a(r)$ are
arbitrary, and we have multiplied $c$ by unit time. We require that the
local geometry of ${\cal M}_k$ reduce to the geometry of special relativity.

For a K\"ahler-torus space-time the phase space manifold of topological
instantons, ${\cal M}_k$, was shown by Kobayashi to be symplectic K\"ahler
\cite{kob}. This means that the holonomy group of ${\cal M}_k$ is $%
Sp(1)\simeq SU(2)$, and immediately implies that the Ricci tensor vanishes.
Since ${\cal M}_k$ is of real dimension four, a Ricci-flat K\"ahler manifold
is in fact hyper-K\"ahler. Hyper-K\"ahler metrics on four-manifolds have
been studied in depth since they have been found to be important to both the
gravitational instanton and the BPS magnetic monopole \cite{gibbons}\cite
{atiyah}. Our goal is now to determine the possible hyper-K\"ahler metrics
on ${\cal M}_k$.

Assume that the hyperK\"ahler metric on ${\cal M}_k$ is complete and
non-singular on an open set of the topological instanton phase space, $%
U\subset {\cal M}_k$. Assume local isotropy of the universe, so that the
phase space ${\cal M}_k$ admits $SO(3)$ as a group of local isometries. Let
us also assume that on $U$ the orbits defined by the action under the
isometries are three-dimensional. Then, the only complete, non-singular, $%
SO(3)$-invariant hyperK\"ahler metrics on four-manifolds with
three-dimensional orbits are: the flat metric, the Atiyah-Hitchin metric,
the Taub-NUT metric with positive mass, and the Eguchi-Hanson metric \cite
{gibbons}. The Taub-NUT and Eguchi-Hanson metric admit another $U(1)$ so
they are in fact $U(2)$-invariant. Only Taub-NUT and the Eguchi-Hanson
metric appear to be compatible with the massless particle metric (\ref
{phomet}). If we also place on $M_{massless}$ its natural complex structure
and note that it is invariant under the natural $SO(3)$ action, then only
the Eguchi-Hanson metric is compatible with the local geometry of special
relativity. The Eguchi-Hanson metric is of the form%
$$
ds^2=\left[ 1-\left( \frac ar\right) ^4\right] ^{-1}dr^2+\frac{r^2}4\left(
\sigma _1^2+\sigma _2^2+\left( 1-\left( \frac ar\right) ^4\right) \sigma
_3^2\right) ,
$$
where $\{\sigma _i\}$ is the dual basis for $so\left( 3\right) $. In terms
of Euler angles, we define
$$
\begin{array}{ccl}
\sigma _1 & = & d\varphi \sin \theta \cos \psi -d\theta \sin \psi , \\
\sigma _2 & = & d\varphi \sin \theta \sin \psi +d\theta \cos \psi , \\
\sigma _3 & = & d\varphi \cos \theta +d\psi .
\end{array}
$$
In Euler coordinates the Eguchi-Hanson metric becomes
\begin{equation}
\label{metric}ds^2=\left[ \gamma (r)\right] ^{-1}dr^2+\frac{r^2}4\gamma
(r)(d\psi +\cos \theta \ d\varphi )^2+\frac{r^2}4(d\theta ^2+\sin {}^2\theta
\ d\varphi ^2),
\end{equation}
where $\gamma (r)\equiv 1-(a/r)^4$. Compare (\ref{metric}) with (\ref{phomet}%
). Note that while the metrics are similar, the two-sphere in the
Eguchi-Hanson metric is a function of $r$. With this observation our
physical parameterization of the open set follows.

The Euler angles $(\theta ,\varphi ,\psi )$ in the Eguchi-Hanson metric (\ref
{metric}) define the direction of the propagation, and the radius of the
sphere is the speed of the massless topological instanton, as we saw for the
massless particle above. For the instanton to carry energy and momentum and
still remain massless, the energy-momentum relation in special relativity
implies that the topological instanton should move at the speed of light.
Thus the speed of light in this theory is also subject to spatial variation.
However, the spatial coordinate $r$ in the tangent space is a little
ambiguous, e.g., where is the origin? We shall suppose that the origin is
Earth, and that $r=2c-\alpha \parallel {\bf d}\parallel $ where $c$ is the
present terrestrial speed of light in vacuum multiplied by unit time, $%
\alpha $ is a dimensionless constant, and $\parallel {\bf d}\parallel $ is
the distance from Earth. This is our physical parameterization of ${\cal M}_k
$.

The speed of light is an experimental constant on length scales much smaller
than cosmological scales, so that we may continue to use the energy-momentum
relation locally. Any variation in the speed of light in vacuum is
presumably on very large spatial scales.

\section{An experiment is proposed}

Let us propose then how one might detect a difference in the speed of
distant light when compared with terrestrial light, $c$.

Assume that the speed of light $\tilde c$ is not an absolute constant when
viewed at very large spatial scales. To measure deviations in speed between
distant light, $\tilde c$, and terrestrial light, $c$, one presumably
examines photons that have interacted with matter in the early universe and
have, since then, traveled unimpeded through space. Some of these photons
eventually enter a detector (unimpeded travel may require that the detector
be space-based). Since it is assumed that no interaction occurs during the
photons' long journey, the energy, $E$, and the linear momentum, $E/\tilde c$%
, are conserved. This implies that the photons travel toward Earth with the
constant speed of light, $\tilde c$, given to them upon emission. The
incoming photons are absorbed by the matter in the detector, and are
re-emitted (photon scattering). We shall assume that the energy and linear
momentum are conserved in photon scattering. By studying the scattered
photons we determine characteristics of the incoming photons. This is the
Compton effect, of course. A derivation is now given for the lowest-order
correction to the Compton formula when the speed of the incident photon is
not the terrestrial speed of light, $c$. For simplicity, we assume that the
photon scatters off a loosely bound electron in the detector (a graphite
detector, for example).

Let the incident photon have energy $E$ and momentum $E/\tilde c$, where $%
\tilde c$ is the incident speed of light. The electron with mass $m$ is
assumed to be initially at rest. After interacting with the electron, the
scattered photon has energy $E^{\prime }$ and momentum $E^{\prime }/c$, and
the electron has momentum ${\bf P}^{\prime }$. Let $c=\tilde c+\Delta c$,
define $\epsilon \equiv \Delta c/\tilde c$, and use conservation of momentum
and conservation of energy to obtain
\begin{equation}
\label{enemom}\epsilon ^2E^2+2\epsilon E(E-E^{\prime }\cos \theta
)+2EE^{\prime }(1-\cos \theta )=2mc^2(E-E^{\prime }).
\end{equation}
We divide (\ref{enemom}) by $EE^{\prime }$ and use $E=h\tilde c/\lambda $, $%
E^{\prime }=hc/\lambda ^{\prime }$ to give
$$
\frac{\epsilon ^2}{1+\epsilon }\frac{\lambda ^{\prime }}\lambda +2\epsilon
\left( \frac 1{1+\epsilon }\frac{\lambda ^{\prime }}\lambda -\cos \theta
\right) +2(1-\cos \theta )=\frac{2mc}h(\lambda ^{\prime }-\lambda
(1+\epsilon )),
$$
where we have used
\begin{equation}
\label{c/c}\frac{\tilde c}c=\frac 1{1+\epsilon }.
\end{equation}
Define $\Delta \lambda =\lambda ^{\prime }-\lambda $ and take the
lowest-order correction to the Compton formula to obtain%
$$
\Delta \lambda \cong \frac h{mc}(1-\cos \theta )+\epsilon \left[ \lambda
+\frac 1\lambda \frac{h^2}{m^2c^2}(1-\cos \theta )\right] .
$$
The second term is dependent on the wavelength, while the terrestrial
Compton effect is obviously independent of the wavelength. Scattering
dependence on the incident wavelength would be a clear signal for spatial
variation in the speed of light.


\begin{thebibliography}{99}
\bibitem{atiyah}  M. Atiyah and N. Hitchin, {\it The Geometry and Dynamics
of Magnetic Monopoles} (Princeton U.P., New Jersey, 1988).

\bibitem{singer}  J-M. Baulieu and I. Singer, {\it Proceedings of conformal
field theory and related topics} (Annecy, France, March 1988).

\bibitem{eguchi}  T. Eguchi and A.J. Hanson, {\it Phys.Lett.} {\bf 74B}
(1978) 249.

\bibitem{nature2}  W.L. Freedman, et al., {\it Nature} {\bf 371} (1994) 757.

\bibitem{gibbons}  G.W. Gibbons, P.J. Ruback, {\it Comm. Math. Phys.} {\bf %
115} (1988) 267.

\bibitem{horowitz}  G.T. Horowitz, {\it Comm. Math. Phys.} {\bf 125} (1989)
46.

\bibitem{kob}  S. Kobayashi, {\it Differential Geometry of Complex Vector
Bundles} (Princeton U.P., New Jersey, 1987).

\bibitem{sandage}  J. Kristian, A.R. Sandage, J.A. Westphal, {\it Astrophys.
J.} {\bf 221} (1978) 383.

\bibitem{lubke}  M. L\"ubke, {\it Math. Ann.} {\bf 260} (1982) 133.

\bibitem{antony}  A. Maciocia, {\it Generalized Fourier-Mukai transforms},
Edinburgh preprint, MS-95-012.

\bibitem{nature1}  M.J. Pierce, et al., {\it Nature} {\bf 371} (1994) 385.

\bibitem{prasad}  M.K. Prasad, {\it Phys. Lett.} {\bf 83B} (1979) 310.

\bibitem{spinrad}  H. Spinrad, S. Djorgovski, I.A.U. Symp. No. 124 in {\it %
Observational Cosmology}, eds. A. Hewitt, G. Burbidge, L.Z. Fang (Reidel,
1987) 129.

\bibitem{nature3}  N.R. Tanvir, T. Shanks, H.C. Ferguson, D.R.T. Robinson,
{\it Nature} {\bf 377} (1995) 27.

\bibitem{wells}  R.O.Wells, {\it Differential Analysis on Complex Manifolds}
(Springer-Verlag, New York, 1980).
\end{thebibliography}
\end{document}